\documentclass[sigconf]{acmart}
\AtBeginDocument{%
  }

\setcopyright{acmlicensed}
\copyrightyear{2025}
\acmYear{2025}
\setcopyright{cc}
\setcctype{by}
\acmConference[WWW Companion '25]{Companion Proceedings of the ACM Web
Conference 2025}{April 28-May 2, 2025}{Sydney, NSW, Australia}
\acmBooktitle{Companion Proceedings of the ACM Web Conference 2025 (WWW
Companion '25), April 28-May 2, 2025, Sydney, NSW, Australia}
\acmDOI{10.1145/3701716.3715561}
\acmISBN{979-8-4007-1331-6/25/04}

\usepackage{array}

\begin{document}

% \tableofcontents

%%
%% The "title" command has an optional parameter,
%% allowing the author to define a "short title" to be used in page headers.
\title{Bridging Culture and Finance: A Multimodal Analysis of Memecoins in the Web3 Ecosystem}

%%
%% The "author" command and its associated commands are used to define
%% the authors and their affiliations.
%% Of note is the shared affiliation of the first two authors, and the
%% "authornote" and "authornotemark" commands
%% used to denote shared contribution to the research.
\author{Hou-Wan Long}
\email{houwanlong@link.cuhk.edu.hk}
\affiliation{%
  \institution{The Chinese University of Hong Kong}
  \city{Shatin}
  \country{Hong Kong}
}

\author{Nga-Man Wong}
\email{s1145167@s.eduhk.hk}
\affiliation{%
  \institution{Education University of Hong Kong}
  \city{Tai Po}
  \country{Hong Kong}
}

\author{Wei Cai}
\email{weicaics@uw.edu}
\affiliation{%
  \institution{University of Washington}
  \city{Tacoma}
  \state{Washington}
  \country{USA}
}

\renewcommand{\shortauthors}{Hou-Wan Long, Nga-Man Wong, and Wei Cai}
%% No italics

%%
%% The abstract is a short summary of the work to be presented in the
%% article.
\begin{abstract}
Memecoins, driven by social media engagement and cultural narratives, have rapidly grown within the Web3 ecosystem. Unlike traditional cryptocurrencies, they are shaped by humor, memes, and community sentiment. This paper introduces the Coin-Meme dataset, an open-source collection of visual, textual, community, and financial data from the Pump.fun platform on the Solana blockchain. We also propose a multimodal framework to analyze memecoins, uncovering patterns in cultural themes, community interaction, and financial behavior. Through clustering, sentiment analysis, and word cloud visualizations, we identify distinct thematic groups centered on humor, animals, and political satire. Additionally, we provide financial insights by analyzing metrics such as Market Entry Time and Market Capitalization, offering a comprehensive view of memecoins as both cultural artifacts and financial instruments within Web3. The Coin-Meme dataset is publicly available at \textbf{\hyperlink{https://github.com/hwlongCUHK/Coin-Meme.git}{https://github.com/hwlongCUHK/Coin-Meme.git}}.
\end{abstract}

%%
%% The code below is generated by the tool at http://dl.acm.org/ccs.cfm.
%% Please copy and paste the code instead of the example below.
%%
\begin{CCSXML}
<ccs2012>
    <concept>
       <concept_id>10002951.10003260</concept_id>
       <concept_desc>Information systems~World Wide Web</concept_desc>
       <concept_significance>500</concept_significance>
       </concept>
   <concept>
       <concept_id>10003120.10003130.10011762</concept_id>
       <concept_desc>Human-centered computing~Empirical studies in collaborative and social computing</concept_desc>
       <concept_significance>500</concept_significance>
       </concept>
 </ccs2012>
\end{CCSXML}

\ccsdesc[500]{Information systems~World Wide Web}
\ccsdesc[500]{Human-centered computing~Empirical studies in collaborative and social computing}

%%
%% Keywords. The author(s) should pick words that accurately describe
%% the work being presented. Separate the keywords with commas.
\keywords{Memecoin, Web3, Multimodal Analysis, Blockchain}
%% A "teaser" image appears between the author and affiliation
%% information and the body of the document, and typically spans the
%% page.

% \received{20 February 2007}
% \received[revised]{12 March 2009}
% \received[accepted]{5 June 2009}

%%
%% This command processes the author and affiliation and title
%% information and builds the first part of the formatted document.
\maketitle

\section{Introduction}
The memecoin phenomenon represents a rapidly expanding segment within the Web3 ecosystem, where internet culture and decentralized finance intersect. Unlike Bitcoin \cite{nakamoto2008bitcoin}, which was purposefully designed as a decentralized digital currency, memecoins are primarily fueled by social media engagement, humor, and cultural narratives, often driven by community sentiment rather than intrinsic utility \cite{stencel2023meme}. Memecoins such as Dogecoin, Shiba Inu, and Pepe Coin epitomize this trend, with Dogecoin reaching an \$80 billion Market Capitalization in 2021, propelled by viral trends and celebrity endorsements. Platforms like Pump.fun have further democratized the creation of memecoins, enabling users to launch tokens without requiring technical expertise, while also empowering decentralized community-driven governance. By September 2024, Pump.fun had facilitated over 1 million token launches, generating \$100 million in revenue and dominating daily token creation on the Solana blockchain.

\textcolor{black}{Unlike memes, which are primarily cultural and humorous, memecoins introduce a financial dimension that connects internet culture to finance. Despite their significant impact on both cultural and financial spheres, memecoins have been understudied. Most research focuses on their financial aspects, such as market dynamics and price fluctuations, while overlooking the cultural and community-driven forces behind their creation and growth \cite{stencel2023meme, krause2024beyond, li2022dark}. Meanwhile, research on memes only examines their cultural attributes, since memes were initially viewed as non-commercial cultural artifacts, without any inherent financial dimension \cite{molina2020makes}. This paper aims to bridge both gaps by analyzing memecoins through their cultural and financial dimensions, offering a comprehensive view of their role in the Web3 ecosystem.} 

This paper makes the following contributions: \textbf{(i) Open-Source Multimodal Dataset:} We introduce Coin-Meme, the first open-source dataset for studying memecoins in the Web3 space, combining visual (logos), textual (descriptions), community (user comments), and financial (Market Capitalization, Market Entry Time) data to explore the intersection of culture, community, and decentralized finance. \textbf{(ii) Multimodal Analytical Framework:} We propose a framework to analyze memecoins across their visual, textual, community, and financial dimensions, providing a holistic view of how cultural narratives, community engagement, and Web3 technologies influence memecoin adoption and market behavior. \textbf{
(iii) Cultural and Community Insights:} Applying our framework to Coin-Meme, we uncover patterns in memecoin adoption, community sentiment, and thematic trends, offering insights into how decentralized communities and culture shape the financial dynamics of memecoins in Web3.

The paper is structured as follows: Section 2 introduces Pump.fun and reviews related meme analysis. Section 3 describes the Coin-Meme dataset. Section 4 presents the analysis framework. Section 5 discusses the results, including clustering, sentiment, and financial metrics. Section 6 concludes the paper.

\section{Pump.fun Memecoins and Related Work}

Pump.fun is a decentralized platform on the Solana blockchain that simplifies memecoin creation through an intuitive interface. Users can create memecoins by specifying attributes like name, symbol, and supply without technical expertise. Once created, the coin attracts community engagement through comments and feedback, which can enhance visibility or slow growth. If its Market Capitalization reaches 69,000\$, the coin can be listed on Raydium, a decentralized exchange (DEX), allowing for liquidity and broader exposure within the Solana DeFi ecosystem.

The rise of Web2 platforms has made memes a key element of internet culture, shaping public opinion, trends, and humor. The study of memes through multimodal analysis has gained attention, enabling researchers to understand how meaning is constructed and shared across digital platforms. For example, Kiela et al. \cite{kiela2020hateful} proposed a multimodal framework for detecting hate speech in memes, combining visual and textual embeddings for classification. Suryawanshi et al. \cite{suryawanshi2020multimodal} developed the MultiOFF dataset for detecting offensive content in memes by combining image and text modalities, while Xu et al. \cite{xu2022met} introduced the MET-Meme dataset, which incorporates metaphorical features for sentiment analysis. Additionally, Ling et al. analyzes visual elements that distinguish viral image memes from non-viral ones and develops a machine learning model to predict meme virality \cite{ling2021dissecting}.

\textcolor{black}{However, the intersection of memes and cryptocurrency within the Web3 ecosystem has been largely underexplored.} In this study, we build on existing multimodal research to analyze memecoins using a framework that integrates visual, textual, community, and financial data, providing deeper insights into their role within Web3.

\section{Dataset Description}
The Coin-Meme dataset consists of four modalities: Textual Descriptions, Visual Content, Community Interactions, and Financial Data. Textual Descriptions are the narratives associated with each memecoin. Visual Content includes the logo image of memecoin. Community Interactions encompass user comments. Financial Data consists of market-related metrics, including Market Capitalization and the Market Entry Time, which quantifies the speed at which a token gains liquidity and market presence on decentralized exchanges. \textcolor{black}{The dataset includes all 3,751 memecoins that were created between January 2024 and November 2024 and subsequently moved to Raydium, providing a comprehensive overview of the memecoin ecosystem.}

\begin{figure*}
  \includegraphics[width=\textwidth]{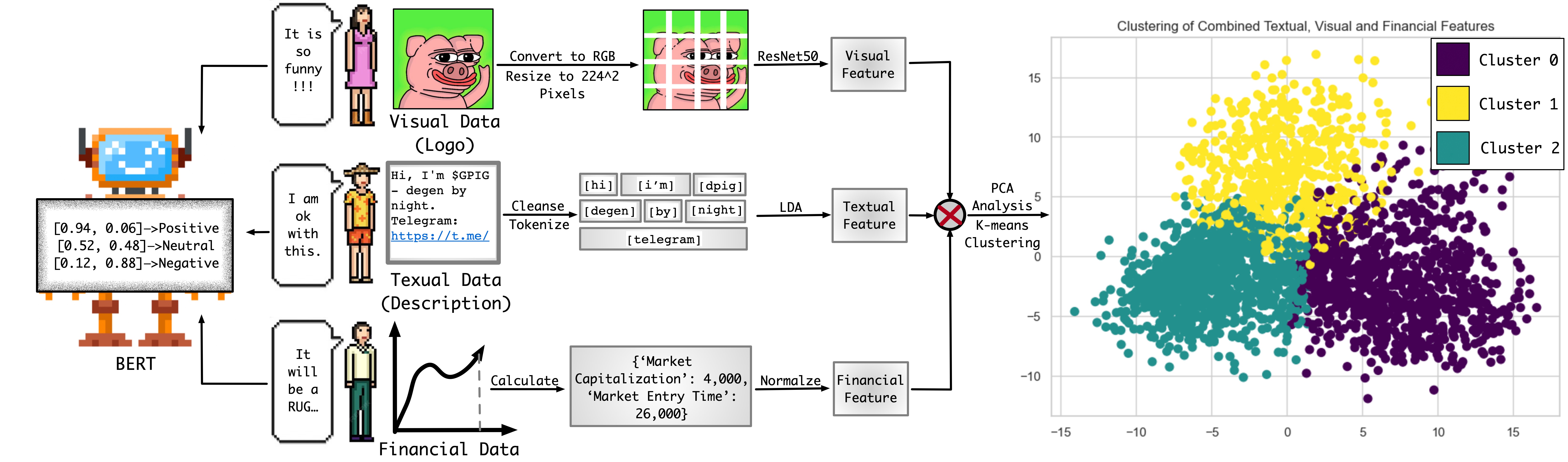}
  \caption{Multimodal clustering of memecoins based on combined textual, visual, and financial feature, with representative descriptions, logos and comments illustrating sentiment categories}
  % \Description{Multimodal clustering of memecoins based on combined textual, visual, and financial feature, with representative descriptions, logos and comments illustrating sentiment categories}
  \label{overview}
\end{figure*}

\subsection{Data Acquisition}
Data acquisition began by extracting token metadata from Dune.com using Pump.fun's program address\footnote{6EF8rrecthR5Dkzon8Nwu78hRvfCKubJ14M5uBEwF6P} including token addresses, creation times, names, and creators. Token-specific URLs were generated by appending token addresses to the base URL\footnote{https://pump.fun/}. A web scraping pipeline built with Selenium and Pandas automated the extraction of textual descriptions, token images, user comments, and financial data.

\subsection{Data Cleansing}
Following data acquisition, the dataset was cleaned to ensure accuracy and consistency. Text data (token descriptions \textcolor{black}{and user comments}) were preprocessed by converting text to lowercase, removing non-alphabetic characters (e.g., URLs), eliminating stopwords, and applying tokenization and lemmatization. Entries with missing descriptions or comments were excluded. Images were resized to 224x224 pixels to match ResNet50 input size, non-RGB images were converted to RGB, and corrupted or missing images were replaced with zero vectors. The time difference between token launch and being moved to Raydium was calculated as Market Entry Time.

\section{Multimodal Analysis Framework}

The multimodal analysis framework presented in this study integrates visual, textual, community interaction and financial data to analyze the cultural and financial dynamics of memecoins.
% \vspace{-0.2em}

\subsection{Feature Representation}
To enable a multimodal representation, textual, visual and financial features were combined into a unified feature space: \textbf{Textual Features:} Latent Dirichlet Allocation (LDA) was applied to tokenize and model topics within the textual descriptions. Each memecoin was represented by a vector of topic distributions, capturing its thematic essence. \textbf{Visual Features:} High-dimensional image vectors extracted using ResNet50 encapsulated visual patterns, including color schemes, shapes, and design motifs. \textcolor{black}{\textbf{Financial Features:} The time difference between the creation and the time used to move the token to Raydium was calculated. Additionally, the Market Capitalization (mcap) of each token was included as a feature, capturing its market value and investment potential. Both quantities were scaled to reflect their relative impact on token adoption and liquidity.}

Textual features, visual features and the financial features were concatenated to form a single high-dimensional representation for each token, effectively capturing its narrative, visual, and financial dynamics.

\subsection{Clustering}
\textcolor{black}{K-Means clustering was used to group memecoins based on their multimodal features due to its efficiency, simplicity, and ability to handle large datasets. The optimal number of clusters was determined using the silhouette score, which evaluates cluster cohesion and separation. K-Means is particularly effective when the number of clusters is predefined, as it minimizes within-cluster variance, making results easy to interpret.} To enhance interpretability and reduce dimensionality, Principal Component Analysis (PCA) was applied before clustering.

\subsection{Analysis}
After clustering, Word Cloud Analysis was conducted on the memecoin descriptions to visualize the frequency of words. Prominent words appear larger in the cloud, helping to identify the dominant themes and keywords associated with each group. This technique uncovers the central narratives and cultural motifs within the memecoin ecosystem. For instance, certain clusters emphasized playful, irreverent themes, while others reflected lighthearted, animal-centric content or politically charged, satirical discourse.

Sentiment Analysis was performed on user comments for each cluster using the \texttt{sentiment-analysis} pipeline from the Transformers library with BERT as the model. The model outputs probabilities for positive and negative sentiments, classifying comments as neutral when both probabilities are close. Comments were categorized into positive, neutral, and negative, revealing sentiment variations across thematic groups and offering insights into emotional tones and cultural dynamics in the memecoin space. Engagement metrics like comments per user were also calculated. Sentiment variability was calculated by calculating the standard deviation of sentiment scores for each coin. Scores were derived from the average of the highest probabilities (positive or negative) across comments. Variability in sentiment helps gauge genuine engagement versus potential bot activity or automated manipulation.

These metrics offer valuable insights into how different themes resonate with users and contribute to the cultural and financial dynamics of the memecoin ecosystem.

% This multimodal framework provides a robust foundation for exploring the diversity and dynamics of memecoins, highlighting the importance of integrating multiple data modalities for cultural analysis. The following sections apply this framework to examine thematic clustering, sentiment analysis, and the broader implications for community behavior.

% \vspace{-0.2em}

\section{Analysis Results on Coin-Meme Dataset}
This section presents the results of applying the proposed multimodal analysis framework on the Coin-Meme dataset. The clustering process identified three distinct memecoin groups as shown in Figure \ref{overview}: Cluster 0 (1,127 memecoins), Cluster 1 (1,535 memecoins), and Cluster 2 (1,089 memecoins).

\begin{figure}[ht]
\includegraphics[width=0.4\textwidth]{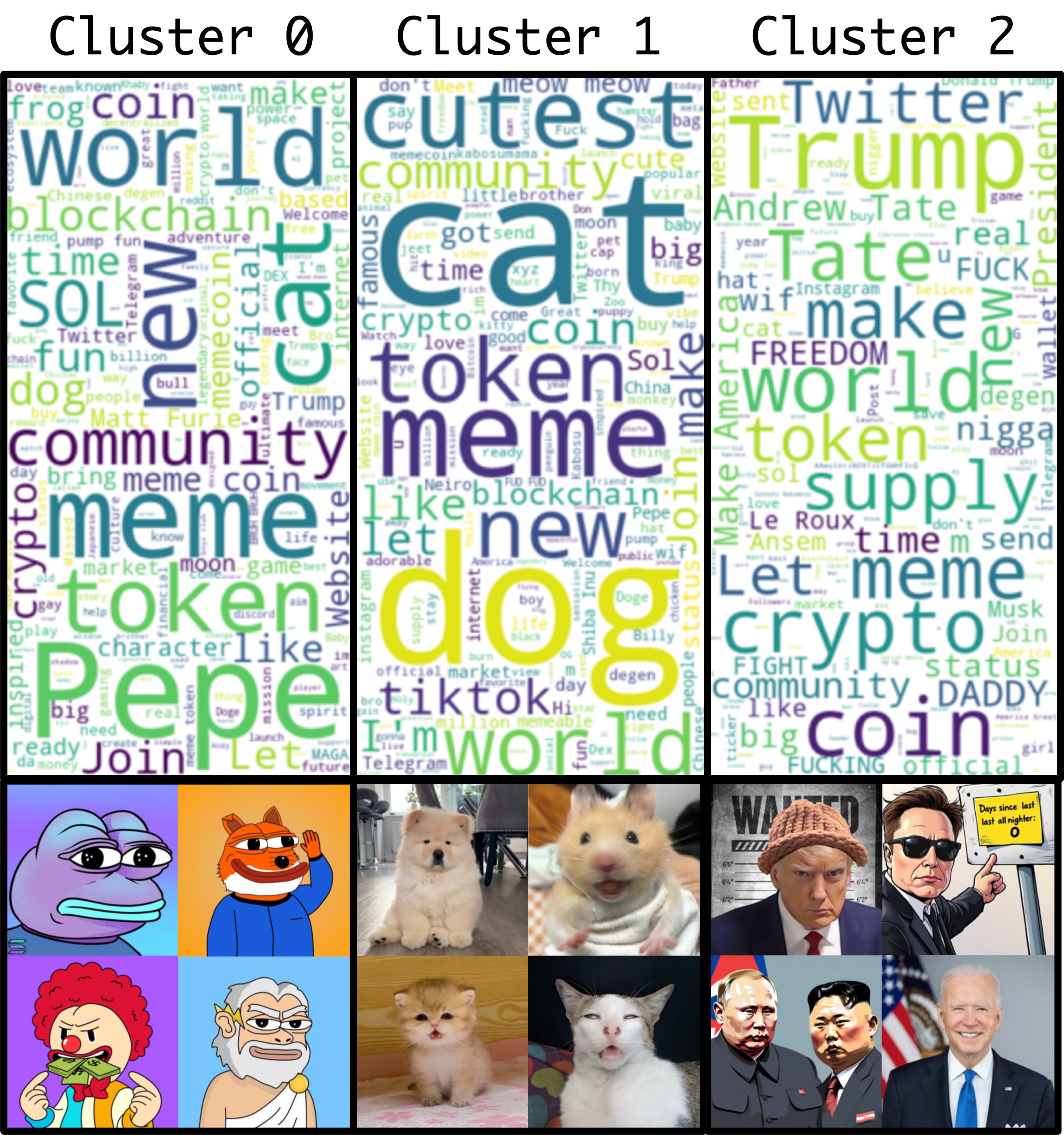}
        \caption{Word Cloud and Representative Images for Each Cluster}
        \label{wc}
\end{figure}

\begin{table*}[t]
\scriptsize
\centering
\begin{tabular}{ccccccc}
\hline
\textbf{Cluster} & \textbf{Positive to Negative Comment Ratio} & \textbf{Comments per User} & \textbf{Mean Sentiment Variability} & \textbf{Mean Market Entry Time (s)} & \textbf{95\% Quantile of MCap (\$)}\\
\hline
0 & 1.23  & 2.97 & 0.3761 & 25,897 &33,557.5\\
\hline
1 & 1.36 & 3.74 & 0.3754 & 7,129 & 24,970.5\\
\hline
2 & 1.64 & 2.78 & 0.3394 &4,728 & 29,508.5\\
\hline
\end{tabular}
\caption{Sentiment, comment and financial statistics per cluster}
\label{stat}
\end{table*}

\subsection{Word Cloud Analysis Results}

Cluster 0 focuses on humor, parody, and internet niche culture. The word cloud (Figure \ref{wc}) highlights terms like "meme" and "pepe," reflecting the irreverent and playful nature of this group. Bold visuals, such as caricatures and fantastical creatures, cater to an audience that values creativity and humor. This cluster exemplifies the humorous and internet culture-driven side of memecoin creation.

Cluster 1 revolves around animal themes and whimsical designs. The word cloud (Figure \ref{wc}) reveals keywords like "cat," "dog," and "cutest," reflecting the lighthearted and relatable nature of the tokens. Descriptions often feature humorous and endearing animal scenarios, while visuals showcase animals like tigers, red pandas, and popular meme animals like Grumpy Cat. The use of words such as "friend" and "fun" underscores the cluster's appeal to users who enjoy playful, emotional content with a whimsical brand identity.

Cluster 2 is characterized by political satire and references to iconic cultural figures, particularly Donald Trump. Descriptions frequently include terms like "Trump," while visuals often feature exaggerated caricatures and hyperbolic portrayals of authority. The word cloud (Figure \ref{wc}) highlights terms such as "trump," "crypto," and "world," emphasizing the cluster's political and engagement-driven nature. This cluster's provocative humor fosters active, often polarizing discussions, making it one of the most controversial and engaging groups in the memecoin space.

\textcolor{black}{Despite the distinct themes of each cluster, several common terms appear across all three, reflecting the shared strategies used by memecoin creators to capture attention. "world" is often used in superlative phrases like "the best in the world," designed to evoke a sense of uniqueness and global appeal. "community" emphasizes the importance of user engagement and collective identity, highlighting the social aspect of memecoins. "meme" is central to the memecoin concept, embodying the cultural and viral nature that drives their popularity. These shared terms demonstrate the common goal of leveraging internet culture and community-building to attract and retain users.}

\subsection{Sentiment Analysis Results}

Cluster 0, focused on humor, parody, and internet niche culture, has the highest positive-to-negative comment ratio (1.23), reflecting the playful and irreverent nature of its content. This suggests a more passive, niche audience with moderate engagement, as indicated by a comments per user rate of 2.97. Cluster 1, centered on animal themes and whimsical designs, demonstrates a positive-to-negative comment ratio slightly higher than cluster 0, highlighting its broader, non-confrontational appeal. The community engagement is notably higher, with 3.74 comments per user, suggesting that its playful, lighthearted nature encourages frequent interactions. Cluster 2, with its focus on political satire and controversial figures, stands out with the highest positive-to-negative comment ratio (1.64), reflecting strong resonance with a passionate, polarized audience. However, its lower comments per user (2.78) suggest less frequent engagement.

\textcolor{black}{The Mean Sentiment Variability values as shown in Table \ref{stat} suggest similar levels of bot activity in clusters 0 and 1, with cluster 2 showing slightly lower variability. This indicates that while all three clusters experience some level of automated influence, cluster 2, with its politically charged content, seems to have less fluctuation in sentiment, possibly reflecting a more controlled or targeted bot activity compared to the other two.}

\subsection{Financial Implication}
\textcolor{black}{We observe distinct financial patterns across clusters, shedding light on the economic behavior of memecoins. The Mean Market Entry Time and 95\% Quantile Market Capitalization were selected to capture key aspects of market adoption, liquidity, and financial stability. The Mean Market Entry Time reflects how quickly a token gains traction on decentralized exchanges, while the 95\% Quantile Market Capitalization reveals the market cap of top-performing tokens, highlighting their scale and potential value.}

\textcolor{black}{Cluster 0, with a Mean Market Entry Time of 25,897 seconds and a 95\% Quantile Market Capitalization of \$33,557.5, shows slower liquidity but higher market cap potential, indicating value due to niche appeal. Cluster 1, with a faster Mean Market Entry Time of 7,129 seconds and a market cap of \$24,970.5, demonstrates quicker adoption and broader appeal but lower market cap. Cluster 2, with the fastest Mean Market Entry Time of 4,728 seconds and a market cap of \$29,508.5, shows rapid entry but potentially lower long-term stability, reflecting the polarizing nature of its content. These metrics provide valuable insights into liquidity, market adoption, and financial stability within each cluster.}

\section{Conclusion}
This study introduces the Coin-Meme dataset and a novel multimodal analytical framework to explore the dynamics of memecoins within the Web3 ecosystem. By integrating visual, textual, community interaction, and financial data, we offer a comprehensive understanding of how memecoins evolve through cultural narratives, community engagement, and market forces in decentralized environments. Our analysis highlights the diversity of memecoin themes, from humorous and playful to politically charged, and uncovers key patterns in user sentiment, community interaction, and potential bot-driven behaviors. Financial implications are also provided. These insights underscore the importance of studying memecoins as both cultural and financial phenomena, bridging internet culture and DeFi. This research not only sheds light on their role within the cryptocurrency market but also offers implications for the future of Web3 communities and governance. Future work could further investigate the impact of social media platforms, sentiment manipulation, and the broader role of memecoins in the evolving Web3 landscape, particularly in relation to DeFi and Web3 governance mechanisms.

\section{Acknowledgments}
This research is supported by Department of Statistics and New Asia College of The Chinese University of Hong Kong.
% Identification of funding sources and other support, and thanks to
% individuals and groups that assisted in the research and the
% preparation of the work should be included in an acknowledgment
% section, which is placed just before the reference section in your
% document.

%%
%% The acknowledgments section is defined using the "acks" environment
%% (and NOT an unnumbered section). This ensures the proper
%% identification of the section in the article metadata, and the
%% consistent spelling of the heading.
% \begin{acks}
% To Robert, for the bagels and explaining CMYK and color spaces.
% \end{acks}

%%
%% The next two lines define the bibliography style to be used, and
%% the bibliography file.
\bibliographystyle{ACM-Reference-Format}
\bibliography{sample-base}

%%
%% If your work has an appendix, this is the place to put it.
% \appendix

\end{document}